# EFFECT OF CARBON-DOPING IN BULK SUPERCONDUCTING MgB$_2$ SAMPLES

M. Paranthaman,[*] J.R. Thompson,[+,¶] and D.K. Christen[+]

[*]Chemical and Analytical Sciences Division, Oak Ridge National Laboratory,
Oak Ridge, TN 37831-6100
[+]Solid State Division, Oak Ridge National Laboratory, Oak Ridge, TN 37831-6061
[¶]Department of Physics, University of Tennessee, Knoxville, TN 37996-1200

**Abstract**

Bulk superconducting samples of MgB$_2$ were prepared by solid state reaction of stoichiometric quantities of Mg turnings and B in a sealed Ta cylinder at 890 ºC for 2 hours. The "as-synthesized" MgB$_2$ samples had a $T_c$ of 39 K, as defined as the onset of diamagnetism. The crystal symmetry was found to be hexagonal with lattice parameters, $a$=3.0856 Å, and $c$=3.5199 Å, similar to the literature values. To study the effect of carbon doping in MgB$_2$, various C-containing samples of x varying from 0 to 1.00 in MgB$_{2-x}$C$_x$ were prepared. Magnetic characterizations indicate that the $T_c$ onset is same for pure and C-doped samples for x = 0.05, and 0.10. However, the shielding signal decreased monotonically with C content, apparently due to the presence of carbon on the grain boundaries that isolates grains and prevents flow of supercurrents on the perimeter.

Keywords: MgB$_2$ superconductors, bulk samples, carbon doping, X-ray studies, Meissner state magnetization

Corresponding Author:
M. Paranthaman
Tel. (865) 574-5045; FAX (865) 574-4961
Email: paranthamanm@ornl.gov

The discovery of superconductivity in $MgB_2$ has generated excitement among the researchers worldwide [1], since $MgB_2$ is a simple binary intermetallic compound with a high transition temperature, $T_c$ of 39 K. In addition, the costs Mg and B are also low, and the crystal structure is unusually simple, consisting of alternating layers of close-packed Mg atoms and "chicken wire" sheets of boron atoms [2]. The structure of a B layer is the same as that of a layer in the graphite structure, with each B located equidistant from three other B atoms (at 1.73 Å), and the next nearest neighbors being a set of six Mg at the vertices of a trigonal prism. In previous work on bulk materials, Larbalestier et al. [3] have demonstrated that these materials are strongly linked, and Bud'ko et al. [4] showed $T_c$ could be increased further to 40.2 K by the boron isotope effect. High temporal stability of supercurrents in bulk $MgB_2$ samples has also been observed in studies of flux creep [5], making the materials potentially attractive for conductor applications. Recently, some reports have focused on producing wires and films for conductor and electronics applications [6-12]; for the latter, physical vapor deposition techniques have been used to deposit $MgB_2$ films on $Al_2O_3$ and Si substrates, followed by either ex-situ [7-9] or in-situ [10-12] anneals.

Ultimately, in order to compete with the high-temperature superconductors in such market, the $T_c$'s will have to be increased. In fact, the $T_c$ of $MgB_2$ was found to decrease linearly at a large rate of $-1.6$ K/GPa under hydrostatic pressure [13]. This suggested that doping Mg with larger cations might increase $T_c$. However, Al [14] and Li [15] doping at the Mg site have been found to decrease the $T_c$. Also, $BeB_2$ is non-superconducting and Be does not substitute for Mg [16]. In further attempts along these lines, we report here our efforts to substitute boron with carbon in bulk $MgB_2$ samples.

The bulk samples of $MgB_{2-x}C_x$ (x = 0.0, 0.05, 0.10, 0.20, 0.50, and 1.00) were prepared in half-gram batches by solid state reaction of stoichiometric quantities of Mg turnings (Aldrich, 98% pure), boron powder, amorphous (Alfa, 99+% pure), and carbon powder (Alfa, 99.9995% pure). The powders were packed in a crimped Ta cylinder that was then introduced into a quartz tube, degassed, and evacuated to $1 \times 10^{-5}$ Torr using a diffusion pump, and sealed. The sealed quartz capsule was introduced into a box furnace, where it was heated slowly to 890 ºC in 3-4 hours, held at 890 ºC for 2 hours and then furnace cooled to room temperature. The quartz tube remained intact during the reaction. The Ta/quartz tube was opened, and the brownish black powder obtained had a very low two-probe resistance of <0.2 Ohms. The powders were ground and characterized for X-ray, and some were pressed into 3 mm dia x 5 cm long pellets and sintered at 890 ºC in a sealed Ta/quartz tube.

Magnetic characterization of the $MgB_{2-x}C_x$ samples was conducted using a SQUID-based magnetometer (Quantum Design MPMS-7). The instrument was equipped with a high homogeneity 7 T magnet and with a sample scan length of 4 cm, maintaining the sample in a highly homogeneous magnetic field. Cylindrical specimens with the apparent density of 60 % were used. The magnetic field was applied parallel to the cylinder axis. The structure of the material was established via X-ray diffraction. A Phillips Model XRG3100 diffractometer with Cu $K_\alpha$ radiation was used to perform the X-ray analysis.

Figure 1 shows the powder X-ray diffraction patterns for $MgB_{2-x}C_x$ samples with x varying from 0 to 1.00. In addition to $MgB_2$ phase, MgO impurities and unreacted C peaks were observed in all the carbon-doped samples. As the carbon content increases, the amount of $MgB_2$ phase decreases. By excluding the impurity peaks, the lattice parameters were calculated for all the samples with a hexagonal symmetry and a space group P6/mmm (191), and the variation of lattice parameters with the carbon content x in $MgB_{2-x}C_x$ are shown in Figure 2. The calculated lattice parameters for the pure $MgB_2$ samples are $a$=3.0856 Å, and $c$=3.5199 Å. This is in agreement with the literature values, and indicates no change in the c-axis lattice parameter values for all the high carbon-doped samples. However, there is a small decrease in the a-axis lattice parameter for the carbon-doped samples. Since we observed a mixed phase for the sample with x = 0.5, the solubility of carbon in boron may be less than 2.5% in $MgB_2$ under our experimental conditions.

The temperature-dependent magnetization of the $MgB_{2-x}C_x$ materials, measured in an applied field of 4 Oe, is shown in Figure 3. The most apparent feature is that all materials have the same onset $T_c$ = 39 K, which is consistent with the above observation that no carbon was soluble in the superconductor. A second noteworthy feature is the differing magnitudes of the signals, for samples with very similar size, measured in the same magnetic field. In this figure, the magnetization is based on the *geometrical* volume of the cylindrical samples, with $M = m/V_{geom}$, which ignores the fact that the materials are only 60 % dense. In fact, when zero-field-cooled (ZFC), the pure $MgB_2$ (x = 0) completely screens the magnetic field from its interior by supporting intergranular surface currents. To interpret the magnetic response here and later, we shall account for demagnetization effect using the relation

$$-4\pi(m/VH)_{true} = \frac{-4\pi(m/VH)_{observed}}{[1-4\pi(m/VH)_{observed}D]} \leq 1$$

where the "observed" susceptibility is based on the applied field and the "true" susceptibility is based on the effective field obtained via the demagnetizing factor *D*. For the dimensions of the short cylindrical samples studied here, we have an effective demagnetizing factor $D \approx 1/3$. Using this relation, we obtain (for x = 0) a "true" susceptibility $-4\pi(m/VH) = 1.0$, corresponding to zero flux density in the interior of the sample; this and other results are tabulated in Table I.

In the pure (x = 0) sample, the surface currents collapse near $T_c$, producing the sharp transition into the normal state as seen in Figure 3. However, the high carbon doped samples with x = 0.05 and 0.1 have a much more rounded transition, as is characteristic of small, isolated superconducting particles [17]. We attribute the isolation to the presence of unreacted carbon between $MgB_2$ grains that prevents effective transport of intergrain supercurrents. Hence a subtle question arises in regard to choosing the appropriate volume basis for interpreting the magnetic response of porous materials. For the doped materials, the magnetization based on the volume of superconductor (calculated from the mass and X-ray density) is more appropriate, for the reason cited. For grains with an isotropic average shape, we have $D \approx 1/3$, leading to the respective shielding values of 0.79 and 0.56 contained in Table I

The Meissner Effect response shown in Figure 3 was obtained by cooling from above $T_c$ in the same 4 Oe field. The "true" Meissner susceptibility $-4\pi(m/VH)$ at $T = 5$ K is shown in Table I; as based on the volume of the superconductor. The fractional flux expulsion is small, 0.15, for the pure sample with x = 0. This can arise from the fact that flux is trapped by voids within the porous cylinder. With the decoupled grains in the doped samples, however, a significant fraction of flux is expelled from grains and from the sample, giving the larger Meissner signals tabulated in Table I. Overall it is clear that the presence of carbon is detrimental to the superconductive properties of $MgB_2$. This is in agreement with the recently reported results on carbon substitution [18,19]. Ahn et al. [18] observed a very broad resistive transition with the $T_c$ onset of 41 K and $T_c$ zero of 33 K for $MgB_{1.8}C_{0.2}$ (x = 0.2) as compared to a $T_c$ zero of 38.8 K for the pure $MgB_2$ sample. Takenobu et al. [19] observed in $MgB_2$ samples doped with low carbon contents, a small drop in both a-axis lattice parameter and $T_c$, for samples up to x = 0.06 in $MgB_{2-x}C_x$. However, they observed the onset of phase separation for samples with x = 0.1.

In summary, we prepared carbon-doped samples in $MgB_{2-x}C_x$ (x = 0.0, 0.05, 0.10, 0.20, 0.50, and 1.00) using the solid state reaction of Mg turnings, B powder and C powder at 890 °C in sealed Ta cylinder. From our experimental results, we conclude that the solubility of carbon in boron is probably less than 2.5 % and the presence of carbon is detrimental to the superconductive properties of $MgB_2$ samples.

**ACKNOWLEDGEMENTS**

This work was supported by the U.S. Department of Energy, Division of Materials Sciences, Office of Science, Office of Power Technologies-Superconductivity Program, Office of Energy Efficiency and Renewable Energy. The research was performed at the Oak Ridge National Laboratory, managed by U.T.-Battelle, LLC for the USDOE under contract DE-AC05-00OR22725.

**Table I.** Meissner State Properties of $MgB_{2-x}C_x$ samples at 5 K: the "true" susceptibility.

| $MgB_{2-x}C_x$ at $T = 5$ K<br>$H = 4$ Oe | x = 0 | x = 0.05 | x = 0.10 |
|---|---|---|---|
| **ZFC (shielding)**<br>$-4\pi(m/VH)$ | 1.04 [a] | 0.79 | 0.56 |
| **FC (Meissner)**<br>$-4\pi(m/VH)$ | 0.15 | 0.55 | 0.43 |

[a] Value based on geometrical volume.

**Figure Captions**

Figure 1.   Powder X-ray diffraction patterns for $MgB_{2-x}C_x$ samples with $x = 0 - 1.00$. $MgB_2$ peaks are indexed with (*hkl*) reflections. MgO impurities, and unreacted C peaks are marked.

Figure 2   Variation of the basal plane (*a*) and inter plane (*c*) lattice parameters as a function of carbon concentration in $MgB_{2-x}C_x$. Impurity peaks were excluded for this calculation. The straight line guides the eye.

Figure 3   Magnetic characterization of the superconducting transitions for 60% dense pellets of $MgB_{2-x}C_x$ for $x = 0$, 0.05, and 0.10. Shown are zero field cooled (ZFC) and field cooled (FC) results for applied field of 4 Oe.

11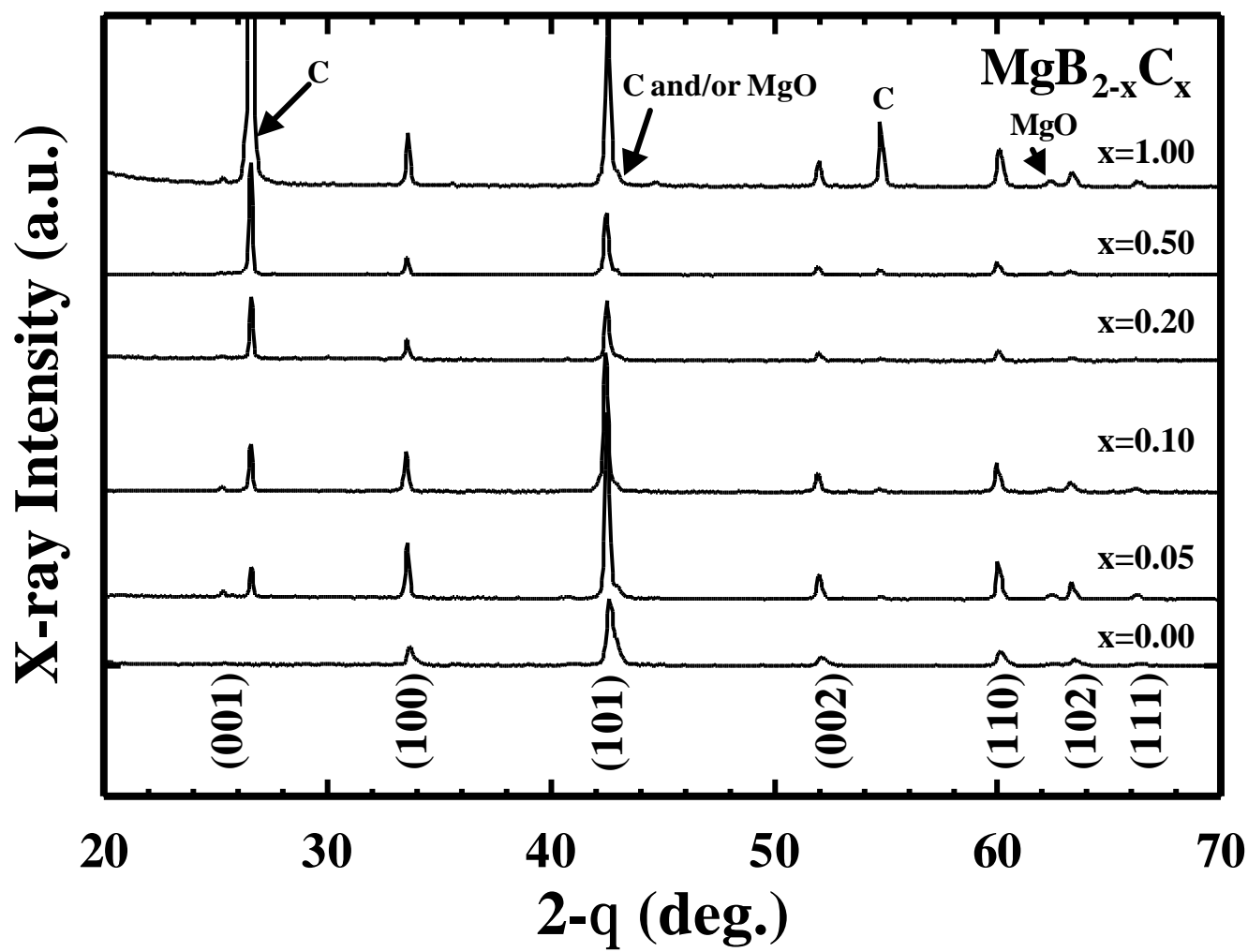

Figure 1 Paranthaman et al.

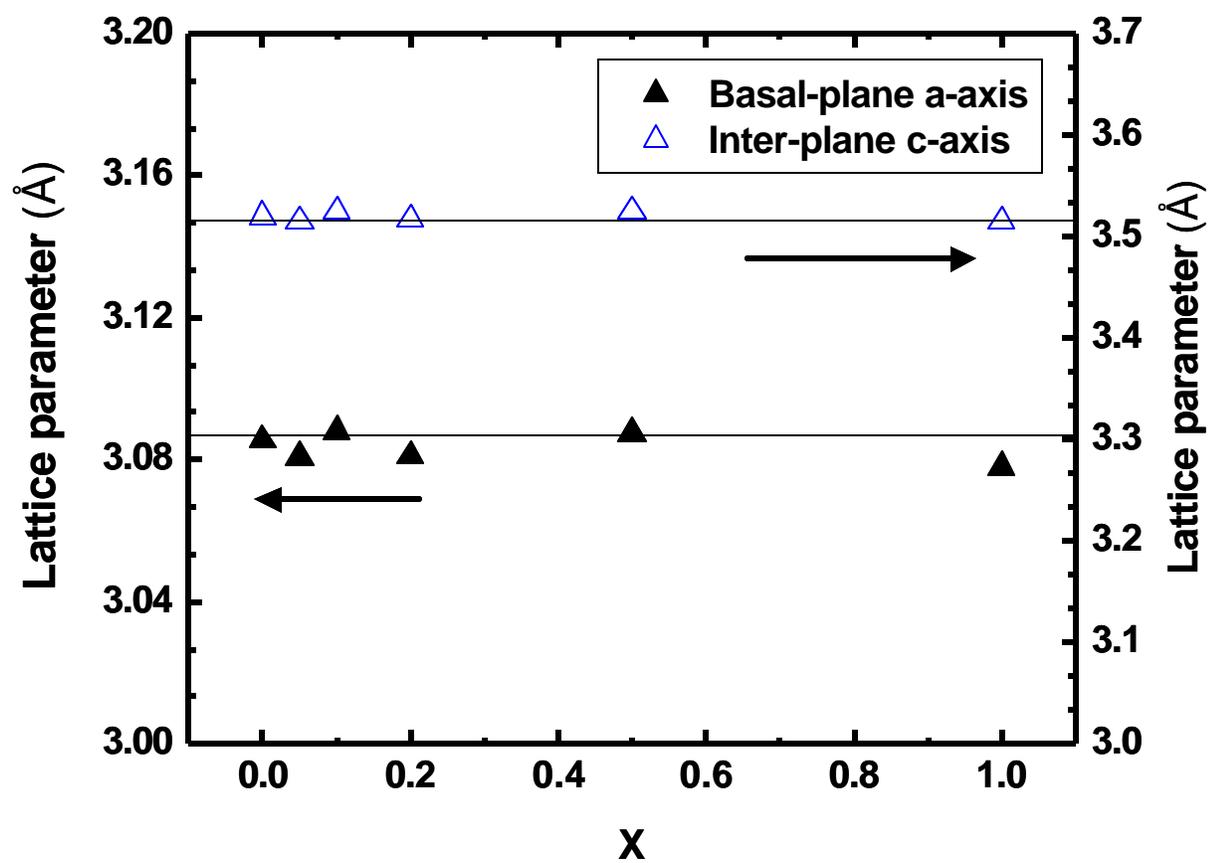

Figure 2 Paranthaman et al.

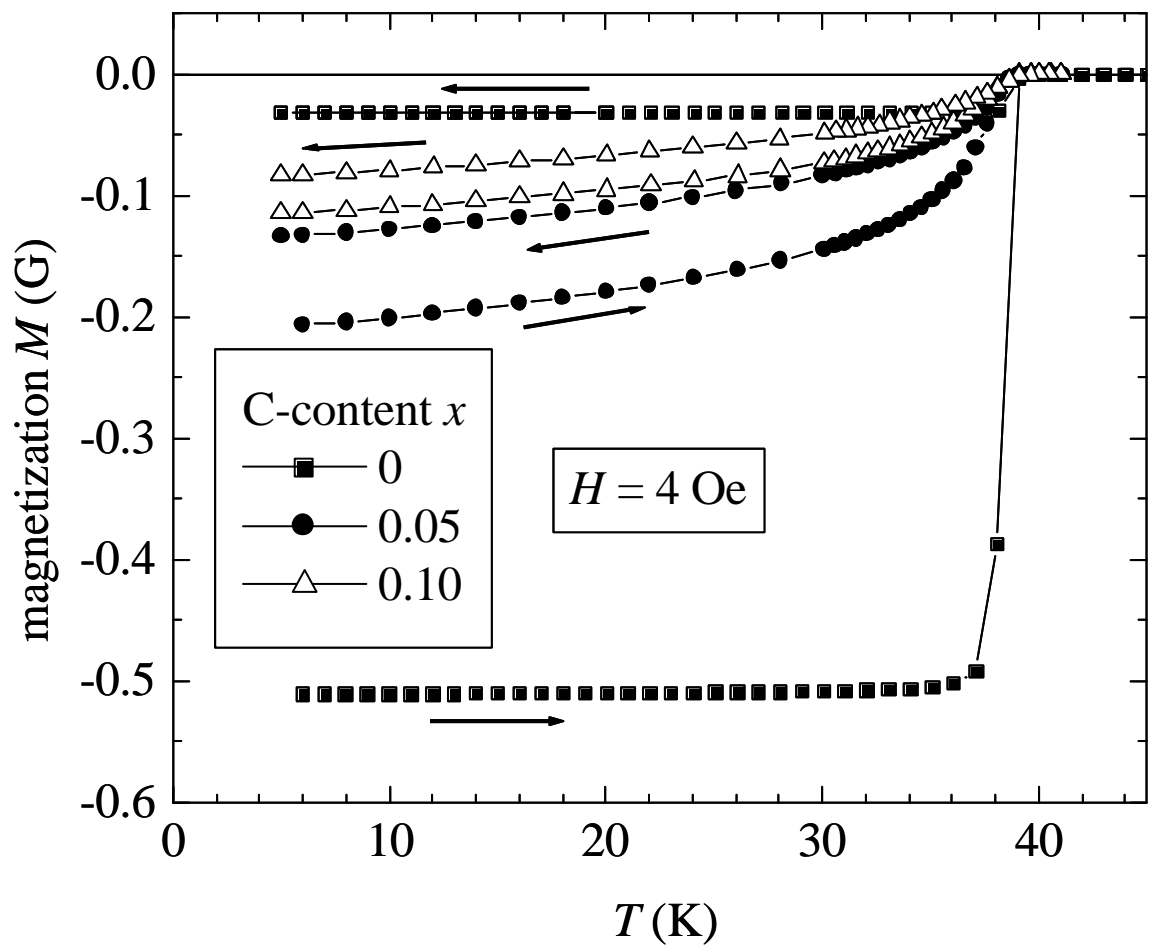

Figure 3 Paranthaman et al.